# Effect of Temperature History During Additive Manufacturing on Crystalline Morphology of Polyether Ether Ketone

Austin Lee[1], Mathew Wynn[1], Liam Quigley[2], Marco Salviato[1,3], Navid Zobeiry[1*]

[1] Materials Science & Engineering Department, University of Washington, Seattle, USA
[2] Mechanical Engineering Department, University of Washington, Seattle, USA
[3] William E. Boeing Department of Aeronautics and Astronautics, University of Washington, Seattle, USA

* Corresponding Author: navidz@uw.edu, 302 Roberts Hall, Box 352120, Seattle, WA 98195, Tel: 1.206.221.3254

## Abstract

Additive manufacturing parameters of high-performance polymers greatly affect the thermal history and consequently quality of the end-part. For fused deposition modeling (FDM), this may include printing speed, filament size, nozzle, and chamber temperatures, as well as build plate temperature. In this study, the effect of thermal convection inside a commercial 3D printer on thermal history and crystalline morphology of polyetheretherketone (PEEK) was investigated using a combined experimental and numerical approach. Using digital scanning calorimetry (DSC) and polarized optical microscopy (POM), crystallinity of PEEK samples was studied as a function of thermal history. In addition, using finite element (FE) simulations of heat transfer, which were calibrated using thermocouple measurements, thermal history of parts during virtual 3D printing was evaluated. By correlating the experimental and numerical results, the effect of printing parameters and convection on thermal history and PEEK crystalline morphology was established. It was found that the high melting temperature of PEEK, results in fast melt cooling rates followed by short annealing times during printing, leading to relatively low degree of crystallinity (DOC) and small crystalline morphology.

## Keywords

Semi-crystalline Thermoplastics, PEEK, Crystallinity, Fused Deposition Modeling, Polarized microscopy, additive manufacturing simulation

## 1. Introduction

Additive manufacturing is a rising technology that enables fabrication of complex geometries with customized requirements. Compared to traditional manufacturing methods, it provides an efficient alternative and often with reduced environmental impacts. In addition, given the flexibility of the technology to fabricate detailed and complex geometries, the potential benefits and broad applications in many engineering fields is apparent [1–4].

Additive manufacturing techniques can employ a wide range of materials including high-performance thermoplastic polymers such as polyetheretherketone (PEEK) with applications ranging from biomedical to aerospace. For polymers, commonly used techniques are selective laser sintering (SLS), stereolithography (SLA), and fused deposition modeling (FDM) [5]. SLS uses polymer powder as feedstock that binds through laser application, SLA uses a photopolymer vat that is cured with a UV laser, and FDM uses polymeric filaments that are melted during printing [6]. Among these processes, FDM is often used due to its relatively cheap and simple process, as well as availability of many types of commercial printers. However, not all commercial FDM printers can process high-performance thermoplastics. Engineering grade polymers like polyetherimide (PEI) and PEEK have high melting temperatures (350-400 °C) and thus require more specialized FDM printers.

PEEK is a semi-crystalline thermoplastic with relatively high transition temperatures, (glass transition temperature, $T_g$, of about 148 °C, and melting temperature, $T_m$, of about 340 °C) and desirable mechanical properties (tensile strength ≈ 100 MPa and Young's Modulus ≈ 3.5 GPa) [7]. PEEK crystalline morphology typically consists of spherulites that exhibit radial symmetry of ordered lamellae with amorphous regions in between. Growth of the lamellae is initiated after a nucleus is formed. The amount of nuclei and the size and distribution of the spherulites impact the mechanical properties of the polymer such as viscoelastic behavior, fracture toughness, and strain-rate sensitivity [8–10]. In general, nuclei are denser than lamellae and perform better at inhibiting crack propagation. Larger spherulites are shown to reduce mechanical properties such as tensile strength and impact resistance of thermoplastics [9].

The rate of nucleation in semi-crystalline polymers – like PEEK – is slow near the melting and glass transition temperatures. The thermal energy near melting temperature leads to high polymer chain mobility which makes it difficult to form a critical nucleus to initiate growth. On the other hand, near the glass transition there is little thermal energy to grow [11]. It follows that there is a parabolic relationship between crystallization rate and undercooling. As the nucleation rate increases, the size of spherulites decreases. The increased number of nuclei quickly impinge on each other during growth leading to a smaller size [12]. As a result, a higher nucleation rate should lead to greater mechanical properties due to the higher amount of denser nuclei [8]. Melt cooling rate has been shown to directly affect the degree of crystallinity (DOC). With faster cooling rates, lower DOC is obtained [13]. For example, while PEEK typically shows a DOC around 30%, at extremely high cooling rates in the range of several thousand °C/min (i.e., several hundred °C/s), the DOC may drop to 10% or lower exhibiting mostly an amorphous morphology [13]. In such a case, reheating above the glass transition temperature will result in cold crystallization and DOC may be increased to around 20% or more [14].

Using FDM technology, a combination of different factors including build plate temperature, ambient temperature, nozzle temperature and printing speed affects the material temperature history, which in turns affects the crystalline morphology and microstructure of the material. While melt cooling rate affects crystalline morphology, melt flow index and surface tension affect the porosity level and microstructure. These in turn control the mechanical properties of printed parts [15,16]. For example, it has been shown that an elevated build plate temperature increases the DOC and hence mechanical properties of parts [17]. At higher nozzle temperatures, viscosity drops leading to lower porosity and increased density. As a result, tensile strength and flexural strength are shown to increase [18,19]. In addition to lowering porosity, higher printing temperatures provide sufficient energy to melt and allow additional time for PEEK crystallization [20]. Other studies found that printing temperature had an interactive effect with ambient temperature, and that the effect of nozzle temperature on tensile strength is not linear [21,22]. One potential reason could be due faster cooling rates at higher nozzle temperatures, leading to lower DOC.

Faster printing speeds decrease the material density due to reduced consolidation, which in turns reduces mechanical properties such as tensile strength. In addition to increasing the density, with slower printing speeds, the filament bonding becomes stronger due to greater inter-filament molecule diffusion, and overall better autohesion. Additionally, surface roughness decreases when the printing speed is increased. It has been shown that nozzle diameters larger than 0.6 mm negatively affect the surface quality when printing speed increases [19,22,23]. In addition to mechanical properties, final part dimensions are also affected by the part temperature history [24,25]. For the build plate, for example, a temperature close to the glass transition temperature of the polymer reduces the residual stresses and dimensional changes in the part. For cases where the thermal difference between the chamber and nozzle temperatures is high, melt cooling rate in the material may vary greatly, leading to low and non-uniform crystallization, in addition to large geometry distortions. In general, a higher ambient temperature should increase part quality and crystallization [22,26].

However, the chamber temperature and other parameters mentioned previously, do not fully explain the heat transfer mechanism occurring in a 3D printer, and consequently observed cooling rates in the material. While conduction occurs between the part and the build plate, as well as internally in the part, convection occurs in the chamber around the part. In general, the airflow pattern and velocity inside the printer control the convective heat transfer coefficient [27]. Aside from natural convection, in commercial printers, forced convection using fans may be used to ensure uniformity of chamber temperature and to accelerate cooling of the nozzle region. The role of the forced convection to maintain proper temperature history becomes more crucial when printing materials such as PEEK with high melting temperatures.

Given the high glass transition and melting temperatures of PEEK, it becomes quite challenging to control the quality of printed parts. Improper control of printing parameters may result in fast melt cooling rates to reduce crystallinity, low melt flow index to increase porosity, and high level of residual stresses leading to excessive distortions in printed parts. Therefore, it becomes imperative to understand and optimize the effect of these parameters on part thermal history.

In this study we aim to explore the effect of chamber convection on part thermal history and evolution of crystalline morphology in PEEK components fabricated using FDM method. A combined experimental and numerical approach is proposed to explore the effect of temperature history. Polarized light microscopy in addition to differential scanning calorimetry (DSC) tests are conducted to characterize the material crystalline morphology. In addition, a combination of experimental thermal profiling using thermocouples, and numerical heat transfer analysis are used to quantify the wide range of melt cooling rates that the material experiences during the printing process.

## 2.        Material and Process

All samples in this study were fabricated using an industrial FDM 3D printer, Funmat HT by Intamsys [28], as shown in Figure 1. The printer has two chamber fans to maintain the chamber temperature around 90 °C during printing. In addition, a nozzle fan with controllable speed is attached near the nozzle with a vent to blow the air directly toward the printing location to control the cooling rate in the material. The vent has an opening of about 5 mm x 15 mm. At maximum speed of 9500 RPM, the fan can blow 4.23 CFM ($\sim 7.18 \; m^3/h$) of air, resulting in an air velocity of about 26.7 m/s. For this study, after initial trials, most printing parameters were kept constant as listed under Table 1. This includes the nozzle temperature of 420 °C, the build plate temperature of 160 °C, and the chamber temperature of 90 °C. The layer build height was about 0.15 mm, and the build orientation was 0°/90° with respect to the edges of the build plate. After the initial trials, the only parameter that was varied in the tests was the nozzle fan speed between 0% -100% full capacity.

The material used in this study is ThermaX PEEK developed by 3DXTECH for 3D printing applications [29]. The nominal glass transition temperature, $T_g$, of this material is 143 °C and the nominal melt temperature, $T_m$, is 343 °C. Several samples were fabricated based on the geometry shown in Figure 2. Upon fabricating samples with different nozzle fan speeds, a correlation between the fan speed and final sample color was observed (Figure 3). This, highlights the effect of fan speed to control the DOC. In general, PEEK with a high degree of crystallinity is opaque light brown whereas amorphous PEEK is translucent amber [30].

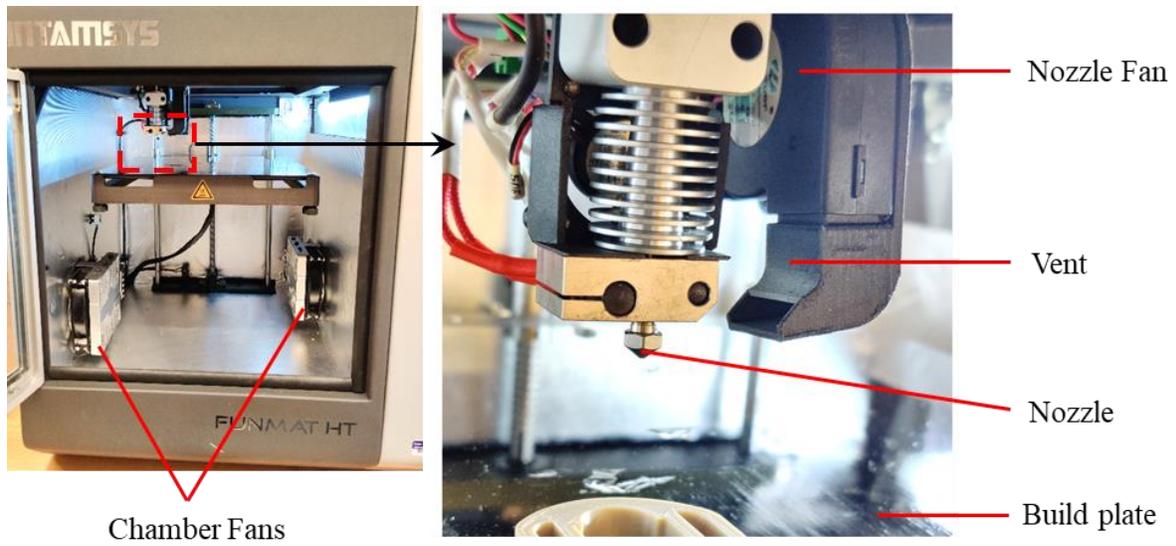

Figure 1. Intamsys Funmat HT printer and nozzle setup.

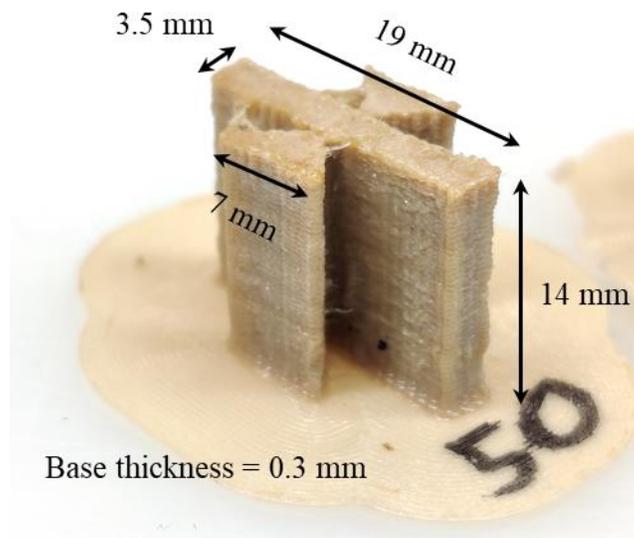

Figure 2. Geometry of Printed PEEK Samples.

Table 1. 3D Printing Build Parameters. All parameters were kept constant except for the nozzle fan speed.

| Process Parameter | Value |
|---|---|
| Print Speed | 12 mm/s |
| Nozzle temperature | 420 °C |
| Build plate temperature | 160 °C |
| Chamber temperature | 90 °C |
| Layer height | 0.15 mm |
| Orientation | 0°/90° |
| Nozzle fan speed | 0% - 100% (0 – 26.7 m/s air velocity) |

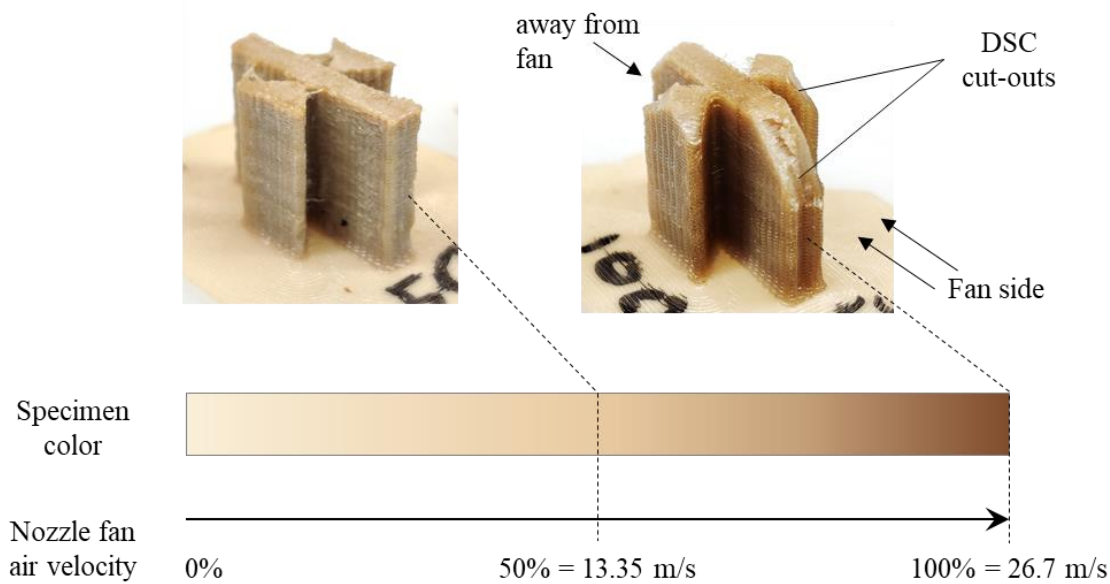

Figure 3. Specimen color as a function of fan speed. The higher the fan speed, the darker the specimen color.

### 3. Characterization of Degree of Crystallinity and Morphology

Upon part fabrication, small samples from different locations were cut to characterize crystallinity using DSC. For example, corners of the parts as shown in Figure 3 for the part with 100% fan velocity were cut for DSC tests. In addition, thin films were prepared and subjected to different thermal histories to investigate the crystalline morphology using polarized microscopy. Results are discussed in the following sections.

3.1 DSC Tests

Thermal analysis measurements were carried out using a Mettler-Toledo DSC 3+ machine under a 50 mL/min nitrogen flow rate. All samples were heated from 25 °C to 425 °C at a rate of 50 or 100 °C/min. The samples were cut from 3 locations on the additive manufactured PEEK parts: the side away from the fan, the side facing the fan, and the base. Several DSC results for samples fabricated with 50% and 100% fan speed are compared in Figure 4. This shows significant cold crystallization for samples fabricated with 100% fan speed. To determine crystallinity, the following equations were used:

$$X_{mc} = \frac{\Delta H_m - \Delta H_g}{H_f^0} \tag{1}$$

$$X_{vc} = \frac{X_{mc}}{X_{mc} + \frac{\rho_c}{\rho_\alpha}(1 - X_{mc})} \tag{2}$$

where $\Delta H_m$ is the enthalpy of melt crystallization during cool-down obtained by taking the integral between specific heat flow rate and a constructed baseline, $\Delta H_g$ is the enthalpy of cold crystallization during heat-up, $H_f^0$ is the enthalpy of fusion for 100% crystalized PEEK (130 J/g [31]), $X_{mc}$ is the mass fraction crystallinity, $X_{vc}$ is the volume fraction crystallinity (referred to as degree of crystallinity in this study), and $\rho_c$ and $\rho_\alpha$ are the densities of crystalline phase (1.4 g/cm$^3$) and amorphous phase respectively (1.26 g/cm$^3$) [31]. Crystallinity results for samples fabricated with different fan speeds are compared in Figure 5 and results are listed in Table 2. Minimum 3 different samples were tested for each condition. There is a clear drop in crystallinity on the side facing the fan (as identified in Figure 3) when the fan speed is increased. The crystallinity of the side away from the fan (as identified in Figure 3) remains relatively the same and hovers around 20%. As the fan speed is increased, the cold crystallization enthalpy increases, while the melt crystallization enthalpy remains relatively constant resulting in lower DOC. With 50% fan speed, the side facing the fan has 16% crystallinity while at 100% fans speed, the crystallinity drops to 11%. After conducting DSC tests with a controlled melt cooling rate of 20 °C/min, a DOC of 26% was obtained as listed in Table 2. For comparison, with 0% nozzle fan speed, an average DOC of 19% was measured. This low DOC compared to the case with controlled cooling rate shows that the material experiences a much faster cooling rates during printing even with 0% fan speed.

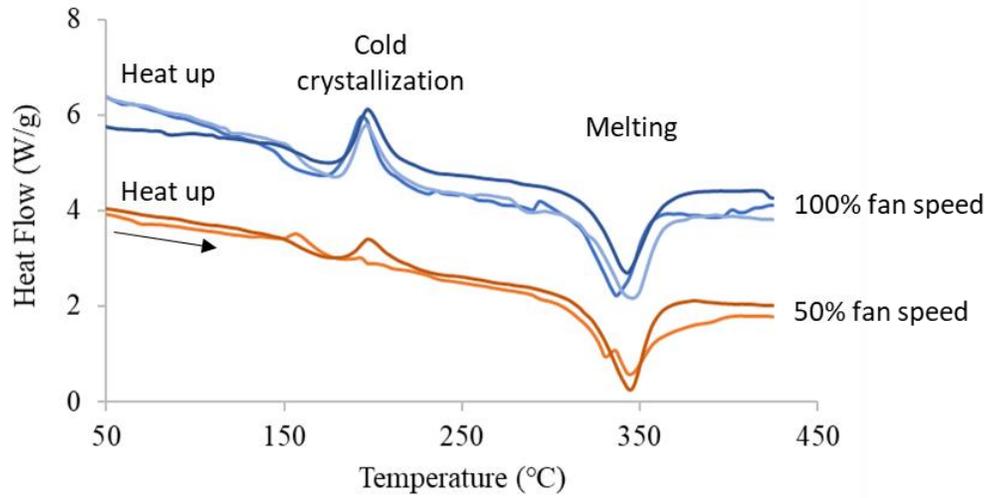

Figure 4. DSC Heat flow as a function of temperature during heating of PEEK samples fabricated with 50% and 100% fan speed in the 3D printer.

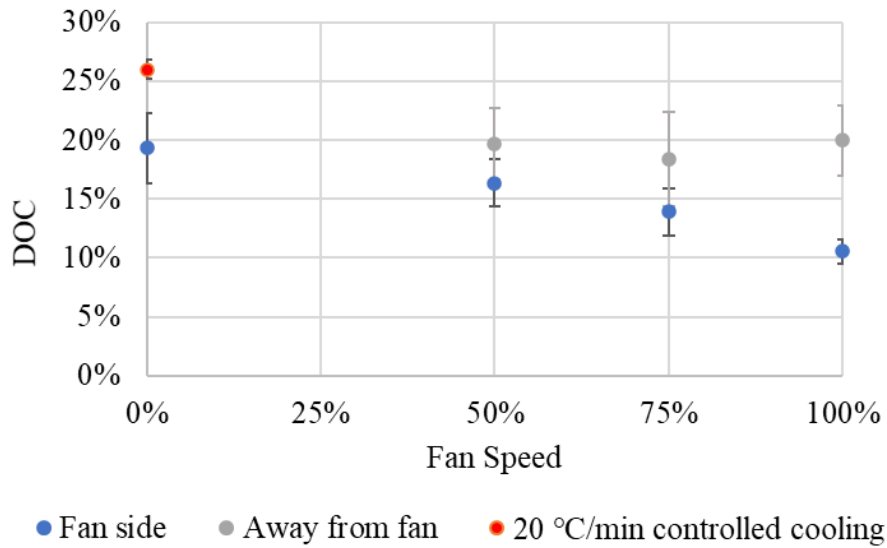

Figure 5. Crystallinity as a function of fan speed and location. Results are compared with DOC for a case with a controlled cooling rate of 20 °C/min.

Table 2. DSC results for degree of crystallinity and enthalpy based on fan speed and DSC sample location

| Sample | $\Delta H_m$ (J/g) | $\Delta H_g$ (J/g) | DOC |
|---|---|---|---|
| 20 °C/min controlled cooling rate | 34.8 | 0 | 26% ± 0.8% |
| 0% fan speed | 27.3 | 0 | 19% ± 3% |
| 50% fan speed – fan side | 26.0 | 2.8 | 16% ± 2% |
| 75% fan speed – fan side | 28.5 | 8.7 | 14% ± 2% |
| 100% fan speed – fan side | 28.0 | 12.9 | 11% ± 1% |
| 50% fan speed – away from fan | 27.8 | 0 | 20% ± 4% |
| 75% fan speed – away from fan | 26.0 | 0 | 18% ± 3% |
| 100% fan speed – away from fan | 27.5 | 0 | 19% ± 0 |

### 3.2  Polarized Microscopy Tests

To investigate the effect of thermal history on crystalline morphology, thin PEEK films were prepared and processed with different thermal cycles before conducting polarized microscopy [32]. Samples of about 1 mm x 1 mm were cut from the additively manufactured parts and placed between two solvent cleaned microscopic glass cover slips. The sample stack-up was then placed onto a compression fixture in a Dynamic Mechanical Analyzer (DMA) 850 from TA Instruments. Samples were heated to 320 °C and held for 20 minutes, and then heated to 400 °C and hold for 9 minutes. During this time, the melted polymer was compressed to a thin film of about 10 μm using a load of 15 N. After compression, the sample was removed and quenched in air quickly to prevent crystal formation. The samples were then reprocessed using a Linkam LTS420 heating stage with controlled cooling to study the effect of cooling rate on PEEK morphology. Finally, the reprocessed samples were imaged using an Olympus BX50 Polarizing Light Microscope (PLM). The analyzer and polarizer were crossed to contrast the birefringent crystalline structures in the sample.

Figure 6 shows results of polarized microscopy performed on a thin sample that was heated to 425 °C, and cooled with a controlled rate of 20 °C/min. The DOC of the sample was previously determined at about 26% using DSC tests (Table 2). This microscopy image shows spherulites with various sizes in the range of approximately 15-30 μm. Given the relatively slow cooling rate, it is expected that the crystals have sufficient time to grow to larger sizes.

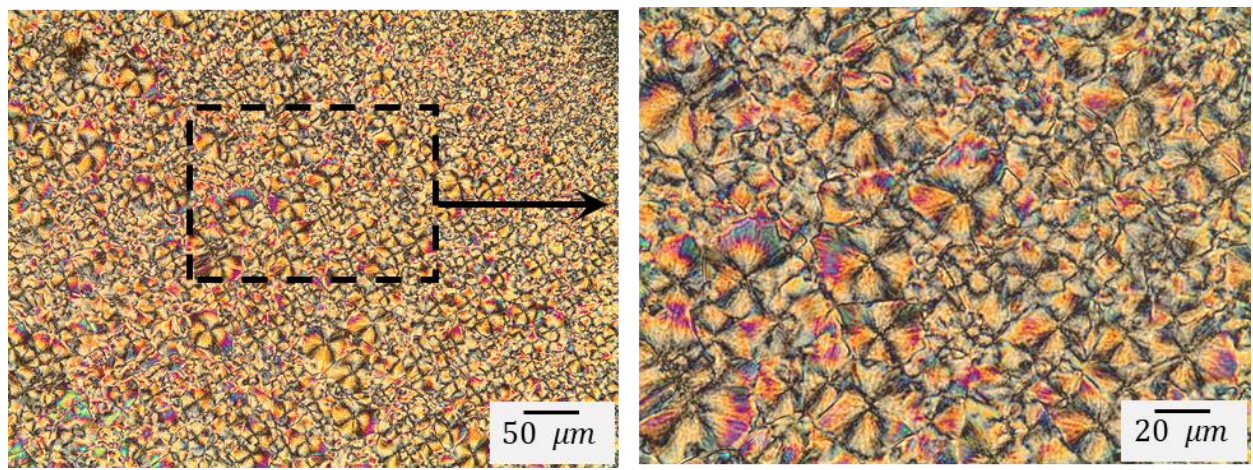

Figure 6. Optical microscope images taken when sample is cooled at 20 °C/min at 20x and 50x magnifications.

To better understand the effect of cooling rate, another thin sample was prepared and subjected to an initial cycle with a fast cool down, followed by a second cycle with a slow cool down as shown in Figure 7a. For the initial cycle, the sample was heated to 420 °C and then air quenched. A fast cool down rate of about 1500 °C/min was estimated using numerical analysis for air quenching of a thin PEEK film sandwiched between two microscope slides. Details of numerical simulations were similar to simulations discussed in the next section.

Polarized light microscopy was conducted on the air quenched sample as shown in Figure 7b. This shows minimal crystallization visible using light microscopy. DSC conducted on the air quenched sample showed a DOC of about 11%. This may be an indication that crystals are extremely small and not visible using light microscopy. Afterward, the sample was heated to 205 °C at a heating rate of 50 °C/min, held for 1 minute, and then quickly removed to recreate cold crystallization. Microscopy showed a small crystal morphology (Figure 7c), and DSC testing showed a DOC of about 22% at this step. This small morphology is an indication that nucleation occurred, but due to limited chain mobility at 205 °C, growth was extremely slow. Finally, the sample was reheated to 420 °C at the rate of 50 °C/min, held for 10 minutes and then cooled at a rate of 20 °C/min. The microscopy image showed large crystal morphology (Figure 7d) and DSC testing showed a DOC of about 26%.

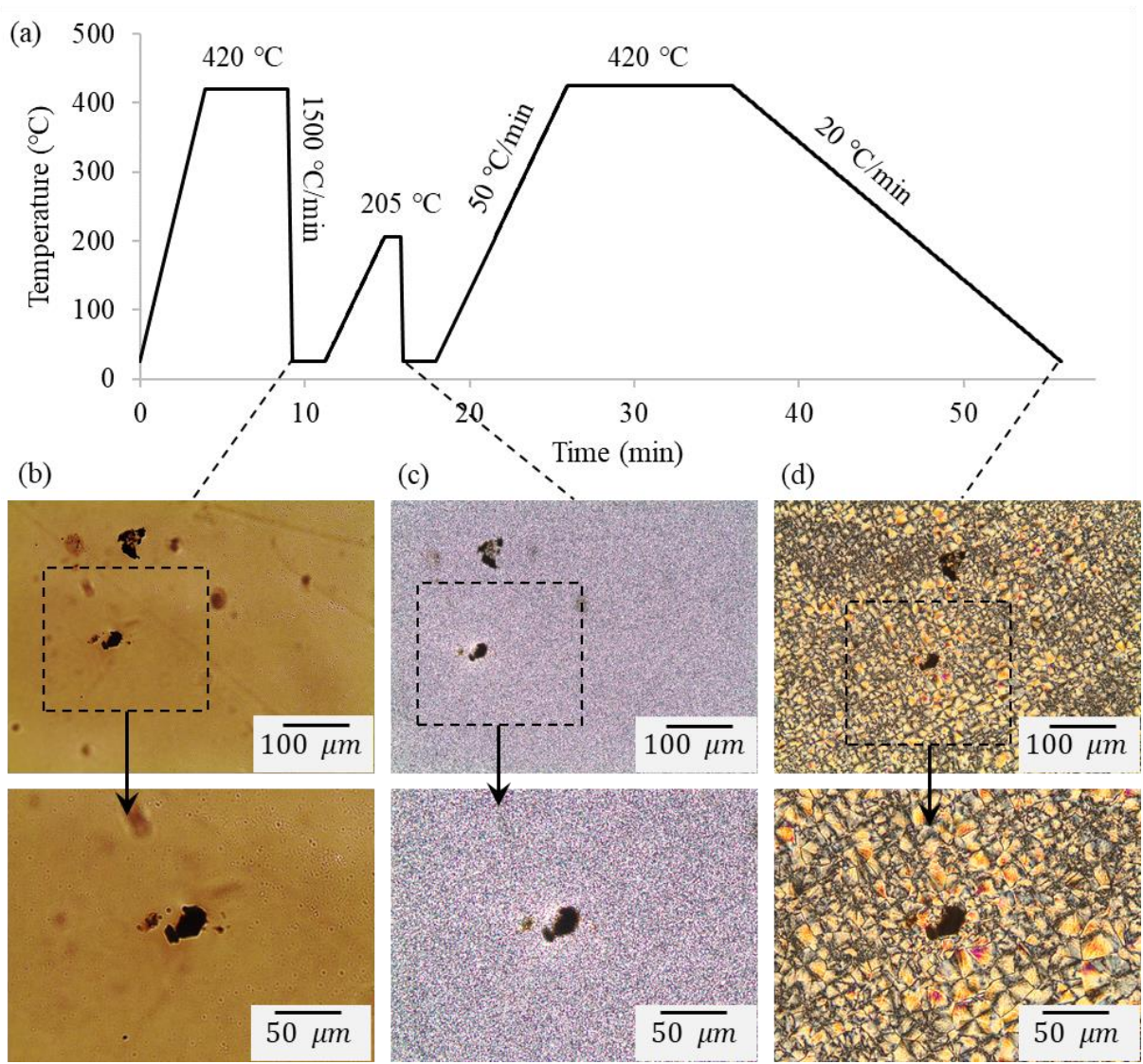

Figure 7. (a) temperature history of a sample initially heated to 425 °C and then air quenched at 1500 °C/min, followed by heating to 205 °C, and finally heating to 420 °C before cooling down at 20 °C/min, (b) crystallinity of air quenched sample, (c) crystallinity after cold crystallization, and (d) crystallinity after melt crystallization.

## 4.    Characterization of the Material Temperature History During 3D Printing

To quantify the range of cooling rates that the material may experience during printing, a combined experimental and numerical approach was implemented. Based on initial measurements using thermocouples (TC) and an IR camera, it was determined that due to extremely fast cooling rates in the 3D printer, an experimental approach alone may not be sufficient to accurately estimate the cooling rates. At the first step, the heat transfer coefficients were measured inside the 3D printer chamber as a function of fan speed. Melted PEEK filaments were printed directly around the tip of a TC with 1 mm diameter, to form a sphere with 3 mm diameter as schematically shown in Figure 8. The chamber temperature was kept at 50 °C. Tests with fan speeds of 0% and 100% were conducted while the cool-down temperatures were recorded by TCs. To estimate heat transfer coefficients (HTCs), finite element (FE) numerical simulation of the heat transfer problem using the commercial finite element code, ABAQUS, was conducted. Material properties for PEEK and the thermocouple alloy used in the simulations are listed under Table 3. Thermal properties from datasheets for PEEK [33] and a nominal thermocouple alloy were used [34]. The crystallization enthalpy was estimated for 20% crystallization based on DSC results in this study. For heat transfer analysis, fully integrated 8-node linear brick elements (DC3D8) with thickness equal to 50 µm were used. Based on different assumptions for HTCs, several simulations were conducted and temperature histories of virtual TCs were recorded and compared with experimental values. Through this trial-and-error, an HTC value of 25 W/m$^2$K was estimated for 0% fan speed, and 200 W/m$^2$K for 100% fan speed. An example of numerical simulation results obtained by ABAQUS/Implicit is shown in Figure 9 for the case with HTC of 25 W/m$^2$K. Temperature distribution in only half of the sample is shown. The numerical and experimental results are compared in Figure 10 for temperature histories of thermocouples subjected to different boundary conditions.

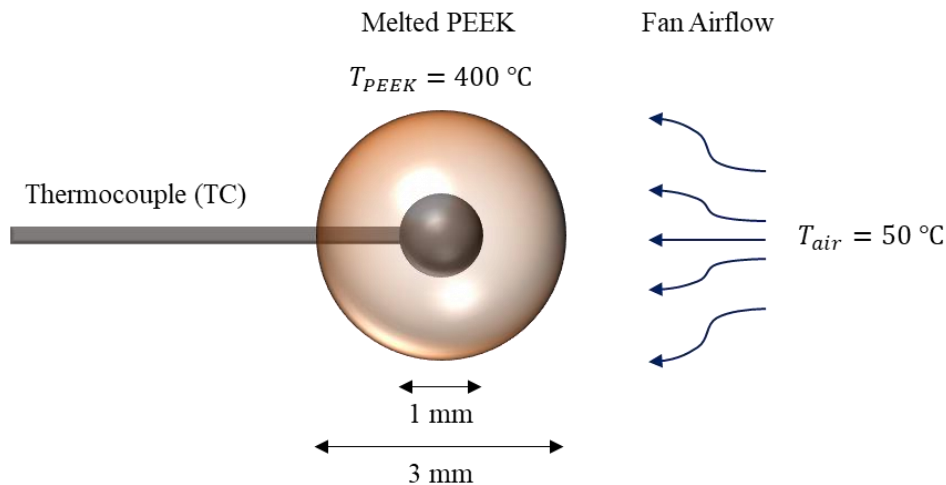

Figure 8. Experimental setup to characterize heat transfer coefficients inside the 3D printer as a function of fan speed.

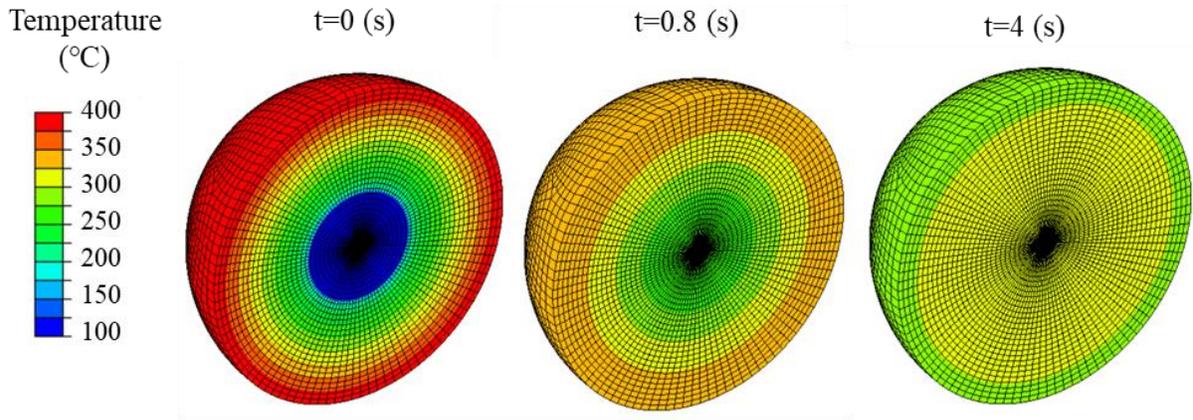

Figure 9. Numerical simulation of cooling of melted PEEK on a thermocouple based on an HTC value of 25 W/m$^2$K.

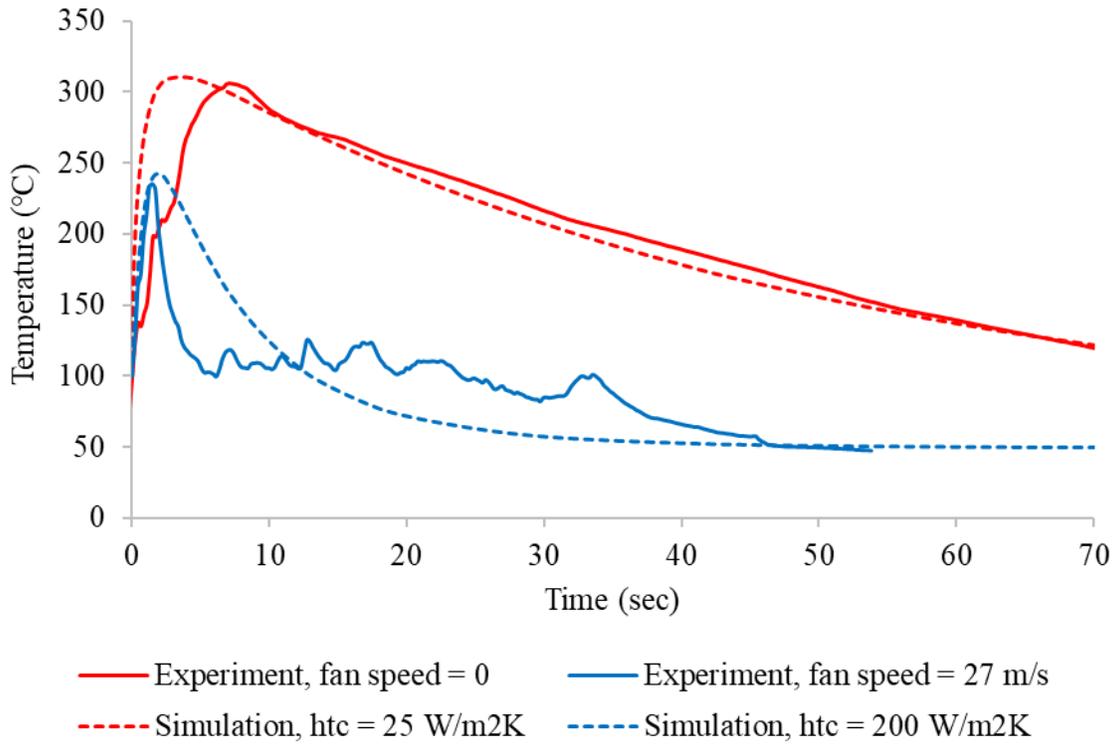

Figure 10. Comparison of TC data with different fan speeds and numerical predictions with different heat transfer coefficients (HTCs).

Table 3. Material properties for PEEK and thermocouple alloy used in numerical simulations.

| Property | PEEK [33] | Sn-Pb Solder Alloy [34] |
|---|---|---|
| Specific Heat Capacity (J/kg) | 1800 | 167 |
| Density (kg/m$^3$) | 1300 | 8400 |
| Thermal conductivity (W/mK) | 0.29 | 50 |
| Crystallization enthalpy (J/g) (DOC =20%) | 27.0 | NA |

While the obtained values for HTCs represent upper and lower bounds of convection around the small PEEK sphere, the HTC around the actual part may vary depending on the location and interaction of airflow with the geometry. From the DSC results (Figure 5), it is obvious that the specimen side facing the fan must experience much larger HTC compared to the side away from the fan. The relative complex geometry of the specimen further enhances the HTC variation. Assuming that the obtained HTC values are upper and lower bounds, we will use them as uniform convective boundary conditions around the sample, to evaluate idealized upper and lower bound thermal responses of the material during printing process. However, it should be noted that the true thermal response of the material may vary between these bounds.

Values obtained for HTCs from the first numerical study were used to setup virtual 3D printings of PEEK samples using FE simulation. Similar to the first study, ABAQUS was used for heat transfer analysis during 3D printing of PEEK samples using material properties provided in Table 3. To speed-up virtual printing simulations, large fully integrated 8-node linear brick elements (DC3D8) with $1.2mm \times 1.2mm \times 0.4mm$ dimensions were used. Based on the printing volume in each step, and printing speed provided in Table 1, printing time for each element was assumed to be 0.2 seconds. Based on printing parameters in Table 1, a build plate temperature of 160 °C and an ambient temperature of 90 °C were considered for virtual 3D printings. For printing, a slightly lower temperature of 400 °C compared to the nozzle temperature of 420 °C was considered. This was to account for the temperature drop from the nozzle to the specimen. In each printing step, convection boundary condition was activated on the surfaces of the newly deposited elements, while the convection was deactivated for the elements covered by the new elements. Printing direction was assumed to be 0 and 90 degrees. Simulation results are compared in Figure 11 for a sample printed with 0% fan speed (Figure 11a, $htc = 25\ W/m^2K$), and a sample printed with 100% fan speed (Figure 11b, $htc = 200\ W/m^2K$). Temperature distributions for three steps of the printing process are compared in Figure 11 for each of these cases. The temperature of the material near the build plate reaches the steady state value of 160 °C for both cases. With 0% fan speed, the sample gradually becomes hotter as printing progresses and the specimen height increases (i.e., away from the build plate). This sample reaches a temperature of about 300 °C at the full height of 14 mm, while with 100% fan speed the sample approaches the ambient temperature of 90 °C as printing progresses.

Based on these simulations, temperature histories are compared in Figure 12 for cases with 0% and 100% fan speed, and for printing locations 2 mm and 10 mm away from the build plate. The temperature histories are taken for the middle of the samples, away from the vertical surfaces. Comparison with temperature histories on the vertical surfaces showed small differences in most cases. Results show that for 0% fan speed, material melt cooling rate is about 5600 °C/min, 2mm away from the build plate, and 2200 °C/min, 10mm away from the build plate. With 100% fan speed, melt cooling rate is almost equal to 12500 °C/min for locations 2mm and 10mm away from the build plate. For all cases, as a new layer is printed over the previous layer, temperature slightly increases. With 0% fan speed, temperature reaches 250 °C and 300 °C for locations 2mm and 10mm away from the build plate respectively. For 100% fan speed, maximum temperature remains below 180 °C. For this sample, the extremely fast-cooling rate followed by a negligible annealing is an indication that crystallinity remains low with 100% fan speed. This is the same conclusion which was made previously using the DSC results (Table 2).

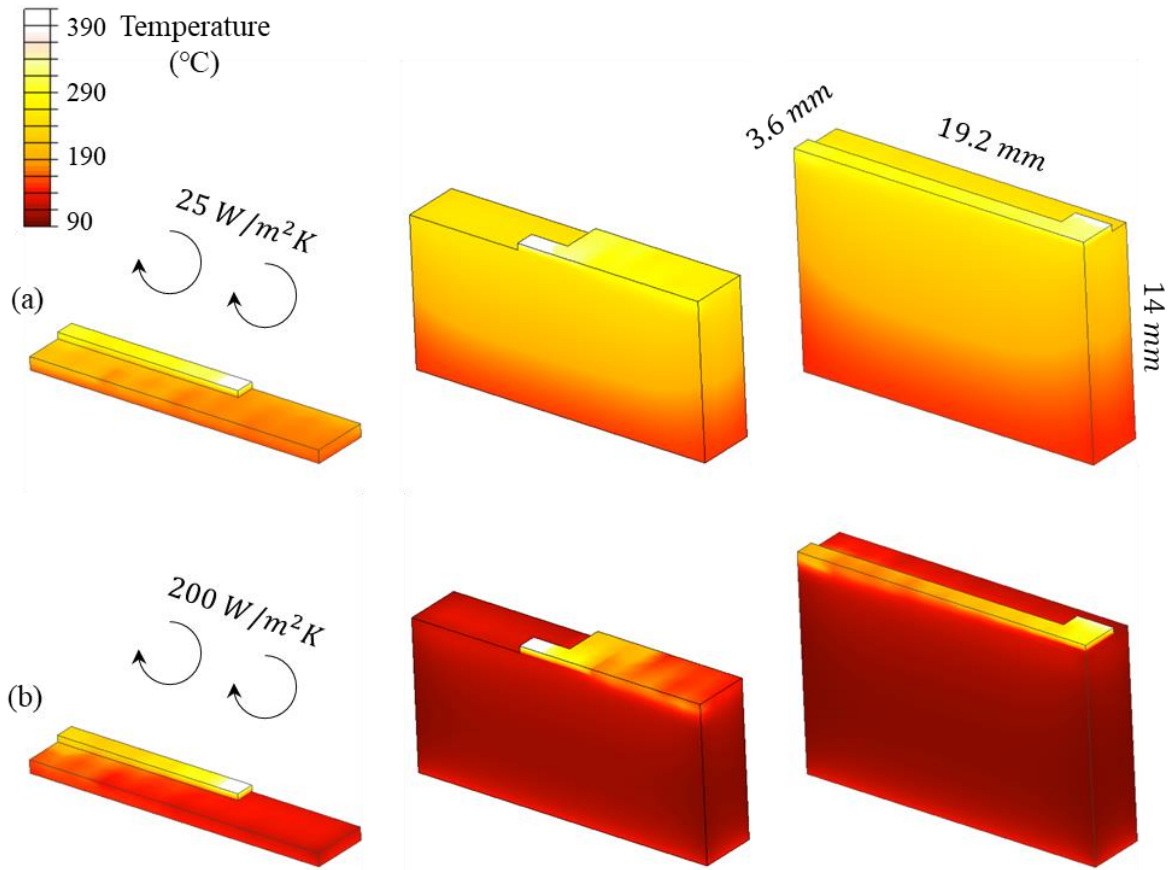

Figure 11. Finite element analysis of heat transfer during 3D printing of two samples with (a) 0% and (b) 100% fan speed (HTC = 25 and 200 W/m2K respectively). (Videos of the simulations are provided in the Supplementary Material).

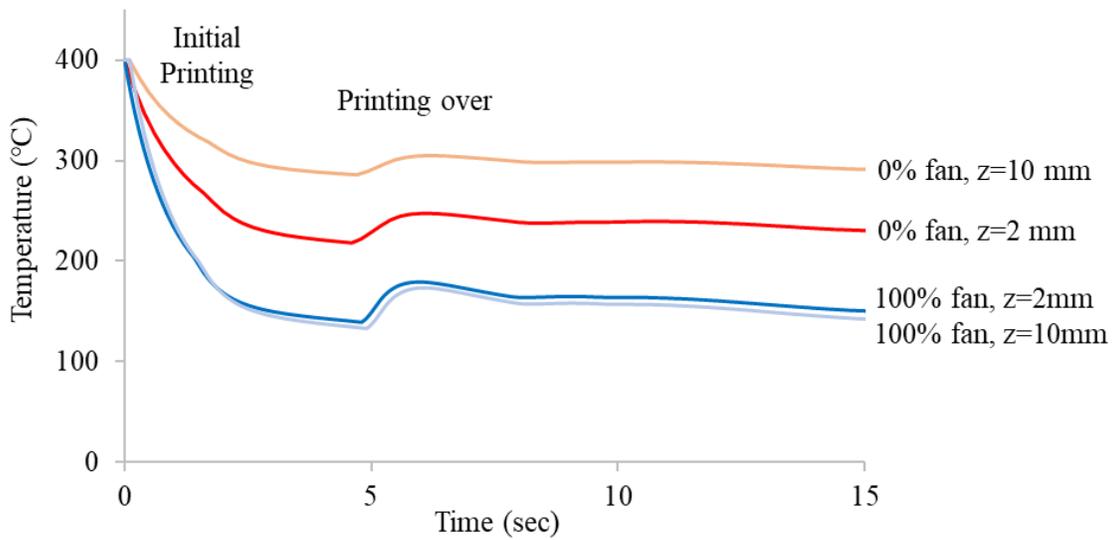

Figure 12. Temperature histories of material during printing based on fan speed, and distance from the build plate, z.

## 5.      Discussion

The temperature history of 3D printed PEEK has a notable effect on its microstructure. From the DSC data seen in Figure 4, Figure 5 and Table 2, the effect of convection controlled by the fan speed can be observed. By increasing the fan speed from 50% to 100% (13.35 to 26.7 m/s air velocity), cold crystallization enthalpy increases. The resulting calculations after melting yields 16% and 11% crystallinity for 50% and 100% fan speed respectively. This relationship is expected as crystallinity is affected by the cooling rate. The low value of crystallinity suggests very fast cooling rates. This was later validated through numerical simulations were cooling rates in the range of several thousand °C/min were estimated.

Microscopy results helped to visualize the effect of thermal history on the material morphology and spherulite structure as shown in Figure 7. Slower cooling rates lead to larger and more distinct spherulites. With the 20 °C/min cooling rate, spherulites were found to measure 15-30 µm in diameter as seen in Figure 6. With very fast cooling rates in the range of thousands °C/min no distinct microstructure was observed as shown in Figure 7b. The DOC of these samples was in the range of 10%. This is an expected DOC from literature for cooling rates over 1000 °C/min [11]. This indicates extremely small spherulite structure not visible with light microscopy. By post processing the parts and increasing the temperature to around 200 °C, small spherites starts to grow further to the point that the structure becomes visible in the polarized microscope as shown in Figure 7c, while increasing DOC to about 20%.

As calculated with numerical simulations, the initial melt cooling rates obtained in the 3D printer are also in the range of thousands °C/min, independent of the fan speed. However, with the slower fan speed, the material goes through a secondary heat cycle while printing over it and remains above 200 °C for about 1-2 minutes. While printing with 100% fan speed, the material does not go through this secondary heat cycle and its temperature mostly remains below 160 °C. This means that the material printed with 0% fan speed should have a similar structure to that shown in Figure 7c, and the material printed with 100% fan speed should have a similar structure to that shown in Figure 7b. This hypothesis can be validated when cross correlating the DSC results obtained for 3D printed samples, and microscopy samples. For example, the sample printed with 0% fan speed, and the microscopy sample that went through a secondary cold crystallization at 205°C, both have a DOC of about 20%. For 100% fan speed, however, the temperature does not reach to the level to cause cold crystallization. This is further validated by DSC results in Figure 4, that shows cold crystallization for samples printed with 100% fan speed.

It should be noted that the cooling rates estimated from virtual 3D printing simulations, can be affected by many parameters including fan speed, build plate temperature, ambient temperature, printing speed, printing direction, part geometry and number of parts printed together. In general, as the delay time between printing layers over each other increases, the part temperature decreases, resulting in faster cooling rates. If printing multiple parts together, or if printing large geometries, the delay time is increased resulting in lower part temperature and faster cooling rates. From the microscopy and DSC tests, we can conclude that this results in cold crystallization of the material.

## 6.     Summary and Conclusions

During additive manufacturing of semi-crystalline thermoplastics, such as FDM processing of PEEK parts, many parameters affect the morphology and microstructure of the material, which in turns affect the mechanical properties and performance of the end-part. In this study, we demonstrated the effect of temperature history of 3D printed PEEK parts on crystallinity and morphology of the material. Using a nozzle fan with variable speed, chamber convection was controlled during printing. Using a combined experimental and numerical approach, the effect of convection on material cooling rate, DOC, and crystalline morphology was investigated. Given the high melting temperature of PEEK, independent of convection, fast melt cooling rates in the range of thousands °C/min were estimated, resulting in negligible DOC of about 10% and extremely small crystal morphology not visible with light microscopy. However, with 0% fan speed, the material goes through a secondary heat cycle while printing over it, to cause cold crystallization. This increases the DOC to about 20%, and the crystal morphology becomes visible in light microscopy. But even at 20% DOC, the crystalline morphology remains distinctively different from the one displayed by materials that are processed with slow cooling rates. With 100% fan speed, the material does not go through this secondary heat step and its DOC remains low. These results highlight the challenges of controlling printing parameters to ensure desirable material morphology while fabricating high performance polymers using FDM process. This study also demonstrates the importance of setting up numerical simulations for heat transfer during printing, to study the effect of different parameters on material thermal history.

**Acknowledgements**

We would like to acknowledge supports by the Boeing Company, Toray Industries, University of Washington, and State of Washington in USA. We would like to thank Boeing company and the Boeing Advanced Research Center at the University of Washington for providing access to the 3D printer, as well as Toray company for providing raw material and access to the heating stage for the microscope. We would also like to thank Materials Science and Engineering department at the University of Washington for providing financial support of the project.